\pdfoutput=1
\documentclass[11pt]{article}

\usepackage[margin=1in]{geometry}
\usepackage[T1]{fontenc}
\usepackage[utf8]{inputenc}
\usepackage{newtxtext,newtxmath}
\usepackage{microtype}
\usepackage{setspace}
\usepackage{parskip}

\usepackage{graphicx}
\usepackage{booktabs}
\usepackage{longtable}
\usepackage{array}
\usepackage{calc}
\usepackage{pdflscape}
\usepackage{caption}
\usepackage{subcaption}
\usepackage[numbers,sort&compress]{natbib}
\usepackage{placeins}
\usepackage{enumitem}
\usepackage{xcolor}
\usepackage[most]{tcolorbox}
\usepackage{url}
\usepackage[hidelinks]{hyperref}

\hypersetup{
 pdftitle={Preventive Care Recommendations by Large Language Models},
 pdfauthor={Eden Avnat, Elia Yanko, Ori Yoran, Raja-Elie E. Abdulnour},
 pdfsubject={Medical AI preprint},
 pdfkeywords={large language models, preventive care, prioritization, time pressure, clinical decision support}
}

\setlist{topsep=0.35em, itemsep=0.2em, parsep=0pt}

\usepackage{titlesec}
\titleformat{\section}{\large\bfseries}{\thesection.}{0.6em}{}
\titleformat{\subsection}{\normalsize\bfseries}{\thesubsection.}{0.6em}{}
\titleformat{\subsubsection}{\normalsize\itshape}{\thesubsubsection.}{0.6em}{}
\titlespacing*{\section}{0pt}{1.2em}{0.5em}
\titlespacing*{\subsection}{0pt}{0.9em}{0.35em}

\newcommand{\keywords}[1]{\vspace{0.5em}\noindent\textbf{Keywords:} #1}

\newtcolorbox{promptbox}{
  enhanced,
  breakable,
  colback=gray!5,
  colframe=gray!45,
  boxrule=0.4pt,
  arc=1.5pt,
  left=7pt,
  right=7pt,
  top=6pt,
  bottom=6pt,
  before skip=0.45em,
  after skip=0.85em,
  fontupper=\small
}

\begin{document}

\begin{center}
{\LARGE \bfseries Preventive Care Recommendations by Large Language Models\\[0.6em]}
Eden Avnat, MD, MPH\textsuperscript{1,2}, Elia Yanko, MD\textsuperscript{2}, Ori Yoran, MSc\textsuperscript{3}, Raja-Elie E. Abdulnour, MD\textsuperscript{4,*}\\[0.6em]
\small
\textsuperscript{1}Gray Faculty of Medicine, Tel Aviv University, Tel Aviv, Israel\\
\textsuperscript{2}Hasharon Hospital, Rabin Medical Center, Petah Tikva, Israel\\
\textsuperscript{3}Blavatnik School of Computer Science and AI, Tel Aviv University, Tel Aviv, Israel\\
\textsuperscript{4}Division of Pulmonary and Critical Care Medicine, Department of Medicine, Brigham and Women's Hospital, Harvard Medical School, Boston, MA, United States\\[0.3em]
\textsuperscript{*}Corresponding author
\end{center}

\begin{abstract}
\noindent\textbf{Objective:} Preventive care services (PCS) extend life, yet physicians often underprioritize highly effective interventions such as lifestyle modifications (Zhang et al., JAMA Network Open 2020). We evaluated whether large language models (LLMs) replicate and augment physician prioritization of PCS under time constraints.

\noindent\textbf{Materials and Methods:} Using Zhang et al.'s validated survey with two patients assessed during long- and short-visit, we compared seven LLMs with historical physicians. We generated 137 simulated physician personas matching cohort demographics and tested three prompts per model. Primary outcomes were concordance with physician rankings (Spearman-correlation) and Consensus-Stratified Agreement (CSA), the proportion of LLM selections rated$\geq$4 that matched physician consensus across extreme ($\leq$25\% or $\geq$75\%) and moderate (25-75\%) agreement strata. Secondary outcomes included life-years gained per prioritized choice (LYGPC), consistency, and selectiveness. Augmentation was assessed by having models revise physician rankings under three progressively informative prompts. $\Delta$LYGPC (model minus physician) quantified the impact of augmentation.

\noindent\textbf{Results:} LLMs closely mirrored physicians (mean Spearman=0.83, SD=0.11), with high CSA at extreme agreement ranges (94\%, 197/210) but low CSA in moderate ranges (21\%, 30/140), where they underprioritized lifestyle services (8.8\% vs. 38\% rated$\geq$4; P$<$0.001). Several models exceeded physicians in LYGPC and consistency (both P$<$0.001) while being more selective. Time constraints affected physicians and LLMs similarly, increasing LYGPC and selectiveness but reducing consistency (P$<$0.001). Augmentation effects varied by model.

\noindent\textbf{Conclusions:} Current LLMs reproduced physicians' time-sensitivity and base-rate prioritization while exacerbating underprioritized lifestyle interventions. Some models improved prioritization performance, but consistent augmentation will require value-aligned training, explicit time-constraint representation, and prospective real-world validation.
\end{abstract}

\keywords{LLM, Preventive-care, Prioritization, Time-pressure, Consistency}

\section{Background}

Preventive care services (PCS) substantially extend life expectancy, yet physicians must make trade-offs during time-limited encounters and competing clinical priorities, often leading to the deprioritization of high-value interventions.\citep{ref1,ref2,ref3} In a survey of 137 primary-care physicians across varied patient scenarios and visit lengths, Zhang et al.\citep{ref1} demonstrated systematic inefficiencies in preventive-service prioritization, with lifestyle counseling receiving disproportionately low attention despite its high expected benefit.\citep{ref1}

Large language models (LLMs) are increasingly discussed as potential decision-support tools because of their accessibility and promising reasoning capabilities.\citep{ref4} However, their performance in PCS prioritization remains unknown.\citep{ref5} This task demands multi-service triage, evidence-aligned ranking, and explicit time management, which are not captured by evaluations that rely on licensing-style examinations or isolated question formats.\citep{ref1,ref3,ref5,ref6,ref7,ref8}

The study by Zhang et al. provides both a value-based benchmark and a historical physician cohort for evaluating decision-support approaches. Building on this foundation, we systematically assessed contemporary LLMs on preventive-service prioritization, examining concordance with physicians, bias propagation or mitigation, internal consistency, time-pressure responses, and life-years optimization. We also tested whether LLMs could improve physicians' PCS prioritization. To our knowledge, this represents the first comprehensive evaluation of LLMs' abilities and augmentation potential in preventive care prioritization.

\section{Materials and Methods}

\subsection{Study Design}

We conducted a cross-sectional simulation study comparing physician survey responses from Zhang et al.\citep{ref1} with seven LLM-generated responses to the identical survey (Figure 1).

\begin{figure}[!htbp]
\centering
\includegraphics[width=0.98\textwidth]{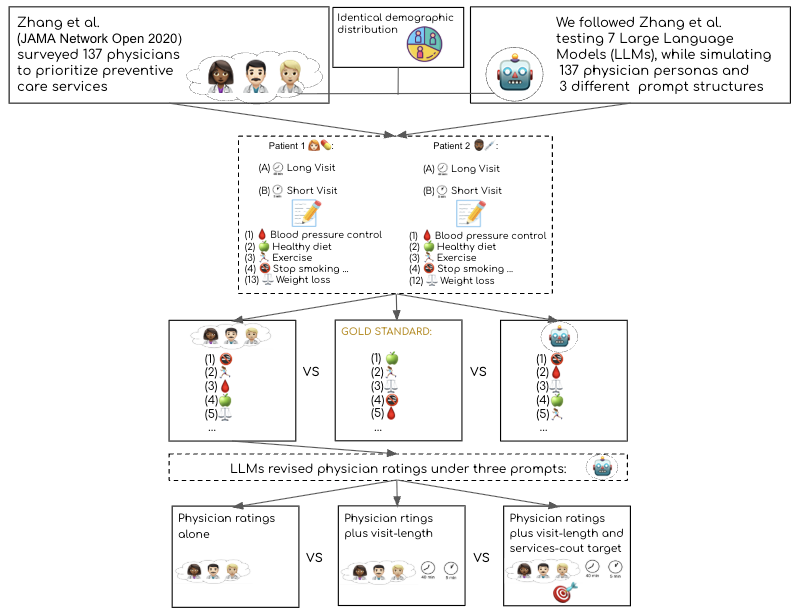}
\caption{Study design: Following Zhang et al.'s methodology (JAMA Network Open 2020), we evaluated seven large language models (LLMs; GPT-5-mini, GPT-o-4-mini, GPT-4.1-mini, Gemini-2.5-flash-preview, Gemini-2.0-flash-live-preview, DeepSeek-R1, DeepSeek-V3-0324) using 137 simulated physician personas and three prompt structures. Physicians and LLMs prioritized preventive services for two patient scenarios under long- and short-visit conditions: a 40-minute visit and a 20-minute acute-visit with 5 minutes available for preventive care, respectively. We compared physicians and LLMs prioritization rating patterns and benchmarked both against life-years-based reference standards from Zhang et al. (GOLD STANDARD). Next, in unrelated sessions, models revised physician ratings under three prompts: (1) physician ratings alone; (2) physician ratings plus visit-length and (3) physician ratings plus visit-length and services-count target. The revised ratings of LLMs were compared with physician ratings.}
\label{fig:study-design}
\end{figure}
\FloatBarrier

\subsection{Survey structure}

Respondents were presented with two hypothetical patient persona: (1) a 50-year-old white woman with hypertension, poorly controlled type 2 diabetes, hyperlipidemia, obesity, current smoking, and family history of breast cancer; and (2) a 45-year-old black man with hypertension, hyperlipidemia, obesity, current smoking, and family history of colorectal cancer.\\ Each patient was evaluated under two encounter lengths: a 40-minute new patient visit (long-scenario) and a 20-minute acute care visit with 5 minutes for preventive care (short-scenario). This 2$\times$2 design yielded four patient-visit combinations (P1L, P1S, P2L, P2S).\\ For each scenario, respondents completed three tasks: (1) rated likelihood of addressing each eligible preventive service on a 5-point Likert scale from "definitely do not discuss" to "definitely discuss"; (2) selected the three most important services; and (3) ranked these services by importance. All questions addressed the current visit, acknowledging that additional services could be addressed during follow-up.

\subsection{Preventive care services and life-years gain}

We evaluated 13 preventive services (patient-1) and 12 services (patient-2) meeting US Preventive Services Task Force grade A/B recommendations using Zhang et al.'s validated framework.\citep{ref1} This estimates individualized life-years gained per intervention based on patient-specific risk factors, with higher-risk patients receiving greater projected benefits (complete methodology provided in the Supplementary File).

\subsection{Physician Data}

Physician data were obtained from Zhang et al.'s 2017 survey (published 2020), which included 137 physicians.\citep{ref1}

\subsection{LLM Selection and Physician Persona Simulation}

We evaluated seven LLMs representing three major vendors, including both reasoning-capable and non-reasoning LLMs: GPT-5-mini-2025-08-07,\citep{ref9} GPT-4o-mini-2025-04-16,\citep{ref9} Gemini-2.5-flash-preview-05-20,\citep{ref10} DeepSeek-R1-0528\citep{ref11} (reasoning-capable); GPT-4.1-mini-2025-04-14,\citep{ref9} Gemini-2.0-flash-live-preview-04-09,\citep{ref10} and DeepSeek-V3-0324\citep{ref12} (non-reasoning).

To mirror the original physician cohort, we generated 137 simulated physician personas using demographic parameters from Zhang et al.\citep{ref1}: sex, age, years in practice, clinical full-time equivalent (FTE), and practice type. Each parameter was sampled from its corresponding distribution in the original dataset, then combined to create complete personas (e.g., a 27-year-old male family medicine physician with less than 5 years' experience, working 60-79\% FTE).

\subsection{System message structure and Survey Prompting}

For each of the 137 simulated physician personas, each LLM was provided with the vendor's default parameters and a standardized system message containing: (1) the physician demographic persona, (2) agreement to participate in a two-part survey, (3) instructions to base all clinical reasoning on guidelines and standards of care in effect as of May 2017, (4) technical instructions for survey completion, and (5) an explanation of the survey structure identical to that used in Zhang et al.'s study.

To assess whether system message formatting influenced model outputs, each LLM was tested under three system message structures:

\begin{enumerate}
\item Single-paragraph version: All information presented in a single continuous paragraph.
\item Bullet version: Information presented in bullet points organized under three headings- Context, Instructions, and Survey.
\item Fusion version: A hybrid format presenting the Context and Survey sections in paragraph form, with Instructions presented as bullet points.
\end{enumerate}

The survey questions were then presented as separate user messages, in the same sequence and wording as in the original physician survey. This design resulted in 137$\times$3$\times$7 LLM chats (personas$\times$message structures$\times$models).

Examples of system message structures and user messages are provided in the Supplementary File.

\subsection{Augmenting Physician Prioritization with LLMs}

We ran an augmentation experiment to test whether LLMs can improve physician prioritization. For each of the four patient vignettes, every LLM received the original vignette together with the mean physician prioritization rating and its 95\% CI for each eligible service, then produced revised ratings on the same 1 to 5 scale.

We tested three prompt families per scenario:

\begin{enumerate}[label=(\Alph*)]
\item Physician ratings only, with no visit-length and no target for the number of services rated $\geq$4; (4- likely discuss this visit, 5- definitely discuss this visit)
\item Physician ratings and explicit visit-length (short or long), with no target for the number of services rated $\geq$4;
\item Physician ratings, explicit visit-length and a target for the number of services rated $\geq$4, set to the scenario specific physician count from Zhang et al.
\end{enumerate}

For each LLM, we ran 30 independent sessions for every scenario by prompt family. Sessions were isolated across prompts and scenarios to prevent intra-session bias. All runs used the same system message that instructed the LLM to review and revise physicians' preventive service recommendations, preserved the temporal anchor to May 2017, and enforced the same output rules as in the main experiment. Illustrative system and user messages appear in the Supplementary File.

\subsection{Primary outcomes}

\noindent\textbf{Spearman correlation with physician rankings ($\rho$):} We converted 5-point importance rankings to ordinal service rankings for physicians and LLMs, then calculated Spearman's $\rho$ to quantify how closely LLM ranking patterns aligned with physician decision patterns. Higher values indicate greater similarity.

\noindent\textbf{Consensus-Stratified Agreement (CSA):} To assess whether LLMs reproduced physicians' base-rate selection patterns, we categorized services into extremes ($\leq$25\% or $\geq$75\% of physicians rating a service $\geq$4) and moderate ($>$25\% to $<$75\%) ranges. For each stratum, we calculated the average proportion of LLMs that also prioritized those services (rating $\geq$4), yielding CSA-extreme and CSA-moderate scores. Higher scores indicate closer reproduction of physician base-rate selection.

\subsection{Secondary outcomes}

\noindent\textbf{Life-years gained per prioritized choice (LYGPC):} Using Zhang et al. lifetables,\citep{ref1} we calculated the total life-years gained from services rated $\geq$4, divided by the total number of prioritized choices per group and scenario.

\noindent\textbf{Optimal LYGPC (upper frontier):} For each scenario, we identified the theoretical maximum achievable benefit by selecting the top-K services based on benchmark life-year values, where K equaled the observed number of prioritized services (ratings $\geq$4). This represented the upper bound of optimal selection.

\noindent\textbf{Consistency (inter-response variability):} We calculated within-group standard deviation (SD) of the 1-5 importance rankings across the 137 personas for each group-scenario-service combination, then summarized across services for between-group comparisons.

\noindent\textbf{Selectiveness:} We counted the number of services each respondent rated $\geq$4 in every scenario, providing a measure of how broad or narrow each model's (or physicians') prioritization set was.

\subsection{Statistical Analysis}

Continuous variables were summarized as means and SD for approximately normally distributed data and as medians and interquartile range (IQR) for non-normal data, and were compared using t-tests or Wilcoxon rank-sum tests, respectively. Statistical significance was set at $\alpha$= 0.05. Benjamini-Hochberg (BH) method was applied to adjust for multiple comparisons and control the False Discovery Rate (FDR). Analyses were performed using RStudio (R version 4.2.2).\citep{ref14} As in Zhang et al.,\citep{ref1} when analyzing 5-point Likert responses, we treated 3.00 as neutral and defined values $\leq$ 2.00 as unlikely and $\geq$ 4.00 as likely.

\subsection{Ethics}

This study used fully deidentified, aggregate physician survey data from Zhang et al.,\citep{ref1} previously approved by the Cleveland Clinic Institutional Review Board and conducted according to the Declaration of Helsinki. No new human subjects were recruited. All additional data were generated through publicly available LLMs and do not constitute human subjects research. Institutional review board approval was therefore not required.

\section{Results}

We simulated 137 physician personas across 7 LLMs (4 reasoning-capable and 3 non-reasoning), each tested under 3 prompt structures, yielding 2,877 completed surveys. Because prompt structures produced similar clinical results, results are reported for the fusion prompt (Supplementary File). Demographic distributions were preserved from Zhang et al.: 54\% (74/137) female, 62\% (85/137) younger than 50 years, and balanced between internal medicine (54\% [74/137]) and family medicine (46\% [63/137]) (Supplementary Table S1).

\subsection{Spearman correlation with physicians' rankings}

LLMs strongly mirrored physician service rankings across scenarios (mean $\rho$-Spearman: 0.83 SD=0.11; range: 0.48-0.99) (Figure 2). Alignment was superior in long versus short visits (median: 0.91 versus 0.82, Wilcoxon-P-value=0.02) and in reasoning versus non-reasoning models (0.9 versus 0.79, Wilcoxon-P-value=0.001). All correlations remained significant and strong after FDR ($\geq$0.68), except GPT-4.1-mini in P2S ($\rho$=0.48, P=0.12), indicating broad preservation of physician prioritization patterns (Figure 2).

\begin{figure}[!htbp]
\centering
\includegraphics[width=0.84\textwidth]{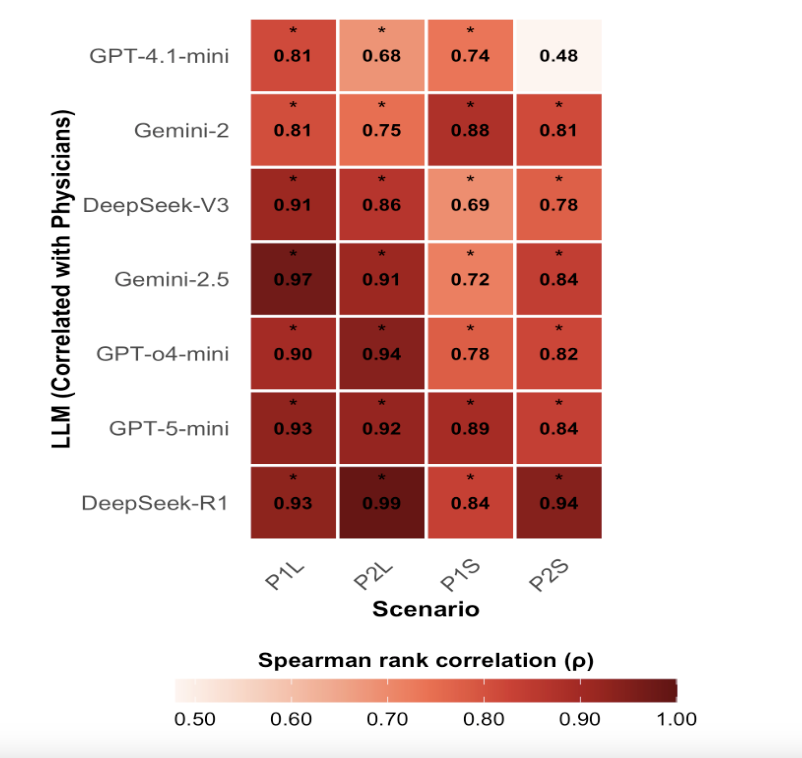}
\caption{Heatmap of Spearman rank correlations ($\rho$) between physicians' and LLMs' service prioritization by scenario. Cell values are $\rho$, with darker red indicating stronger positive correlation. Asterisks denote FDR-adjusted p $<$ 0.05. P1L, P1S, P2L, and P2S denote Patient 1 (Long visit), Patient 1 (Short visit), Patient 2 (Long visit), and Patient 2 (Short visit), respectively.}
\label{fig:spearman}
\end{figure}
\FloatBarrier

\subsection{Consensus-Stratified Agreement (ranking\texorpdfstring{$\geq$}{>=} 4)}

Agreement was consistently high for services with strong physician consensus. CSA at the extremes reached 100\% (63/63), 82\% (40/49), 96\% (54/56), and 95\% (40/42) for P1L, P1S, P2L, and P2S, respectively, yielding an overall rate of 94\% (197/210) (Figure 3).

CSA in the moderate consensus range was low: 21\% (6/28), 21\% (9/42), 14\% (4/28), and 26\% (11/42) respectively (overall 21\%, 30/140 services) (Figure 3). Within this moderate-consensus zone, physicians consistently prioritized lifestyle services more often than LLMs. For relevant services (P1S: diet, exercise; P2S: diet, exercise, weight), 38\% (52/137) of physician personas assigned ratings $\geq$4 compared with only 8.8\% (12/137) among LLM personas (P$<$0.001) (Figure 3; Supplementary Table S2).

\begin{figure}[!htbp]
\centering
\begin{subfigure}{0.98\textwidth}
\caption{Patient 1}
\centering
\includegraphics[width=\textwidth]{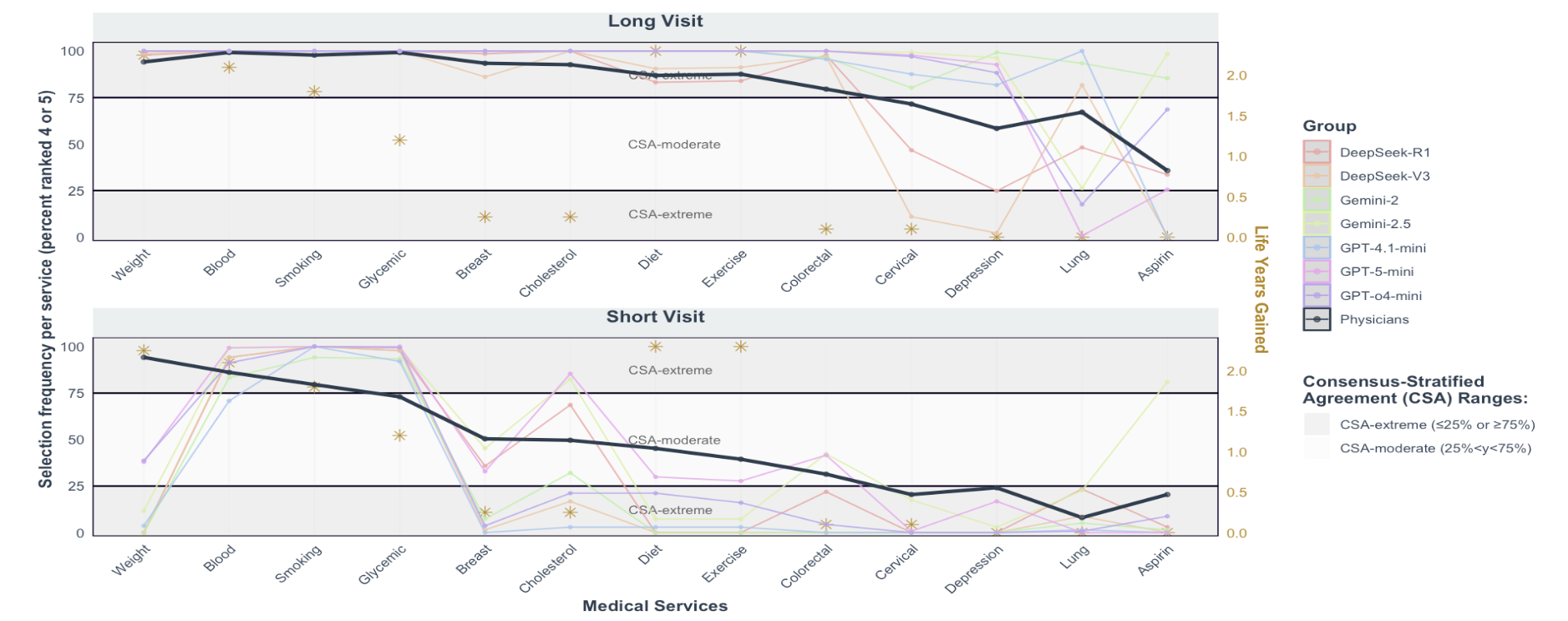}
\end{subfigure}
\vspace{0.4em}
\begin{subfigure}{0.98\textwidth}
\caption{Patient 2}
\centering
\includegraphics[width=\textwidth]{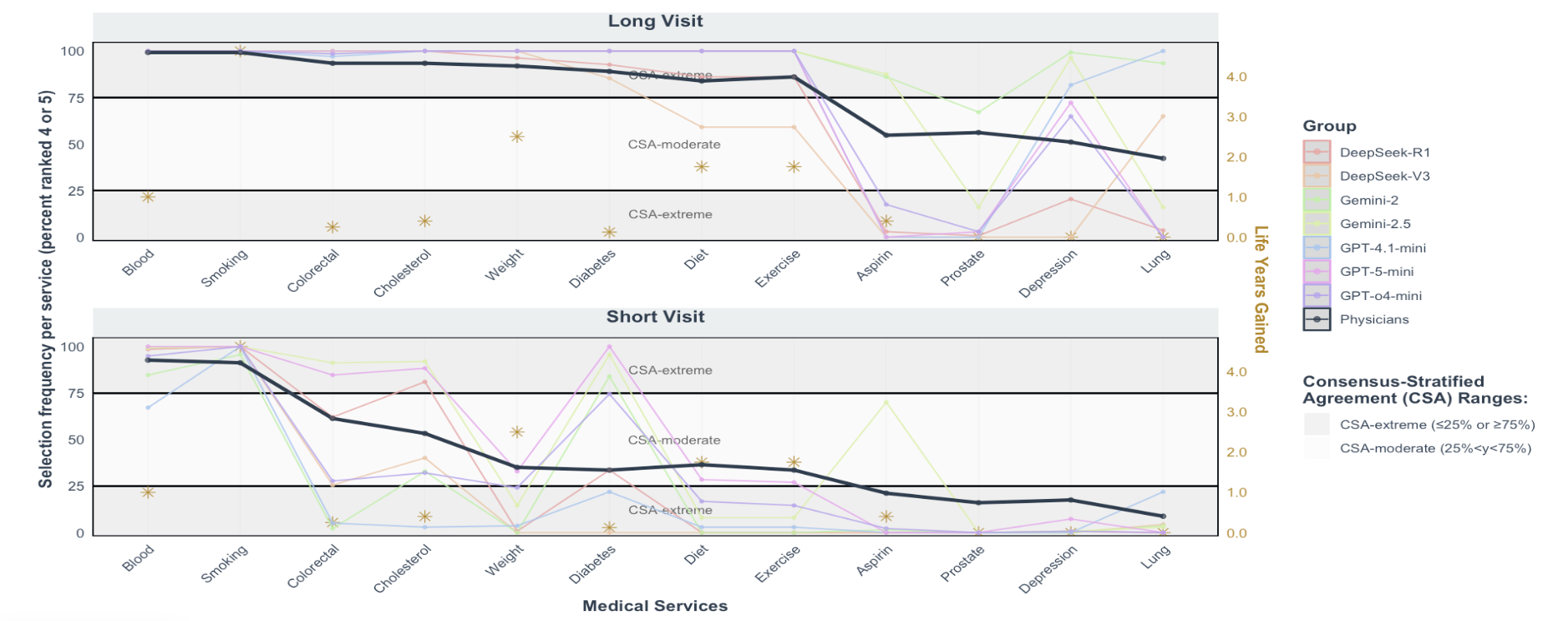}
\end{subfigure}
\caption{Consensus-Stratified Agreement: (A) Patient-1, (B) Patient-2. Each panel includes both visit length (long and short). The black line connects the frequencies of rank $>$4 selections by physicians per service. For example, a frequency of 80\% means that 80\% of physicians ranked a particular service as 4 or 5. Colored lines show LLM selection frequencies. CSA was determined as "extreme" when $>$ 75\% or $<$25\% of physicians or LLMs ranked a service as 4 or 5, indicating strong agreement. Gold asterisks mark benchmark life-years gained per service.}
\label{fig:csa}
\end{figure}
\FloatBarrier

\subsection{LYGPC (ranking\texorpdfstring{$\geq$}{>=} 4)}

Median LYGPC differed across scenarios, ranging from 1.1 [IQR: 1.07-1.14] in P1L, 1.42 [IQR: 1.29-1.59] in P1S, 1.33 [IQR: 1.26-1.42] in P2L, and 1.76 [IQR: 1.52-1.9] in P2S. Several LLMs significantly exceeded physician LYGPC with one outperforming physicians in P1L and four doing so in each remaining scenario (Supplementary Table S3). Notable top performers included GPT-4.1-mini in short visit setting (P1S: 1.67 [IQR: 1.63-1.71]; P2S: 2.43 [IQR: 2.2-2.65]), DeepSeek-V3 in P1L (1.26 [IQR: 1.21-1.31]), and DeepSeek-R1 in P2L (1.5 [IQR: 1.41-1.58]), all FDR-adjusted (Figure 4 A, B; Supplementary Table S3).

No group reached the optimal LYGPC upper frontier in any scenario. Across both patients scenarios, short-visit conditions consistently yielded higher LYGPC than long-visit conditions (Wilcoxon-P-value=0.02). Time-pressure increased LYGPC in 75\% (6/8) of groups for Patient 1 and 63\% (5/8) for Patient 2, while Gemini-2.5 and DeepSeek-R1 showed stable performance regardless of visit length (Figure S1).

\subsection{Consistency}

Across all scenarios, LLMs exhibited substantially lower variability in their ratings than physicians. Median SD was 0.40 [IQR: 0.09-0.55] for P1L, 0.55 [IQR: 0.43-0.79] for P1S, 0.42 [IQR: 0.12-0.6] for P2L, and 0.53 [IQR: 0.44-0.76] for P2S. In contrast, physicians showed significantly higher variation: 0.77 [IQR: 0.59-1.27], 1.39 [IQR: 1.15-1.45], 0.86 [IQR: 0.70-1.44], and 1.33 [IQR: 1.25-1.45] (Supplementary Table S4). Short-visit scenarios increased variability for both patients (Wilcoxon-P-value$<$0.001, both). Time-pressure elevated variability in 75\% (6/8) of groups for Patient 1 and 50\% (4/8) for Patient 2. (Figure S2 A, B; Supplementary Table S4).

\subsection{Selectiveness}

Median selectiveness was 11.5 [IQR: 10.75-12], 4 [IQR: 3-5.25], 9 [IQR: 8.75-10], and 4 [IQR: 3-4.25] for P1L, P1S, P2L, and P2S respectively. DeepSeek-V3 showed lowest selectiveness: 10 [IQR: 9-10], 3 [IQR: 3-3], 8 [IQR: 7-9], and 3 [IQR: 2-3] across scenarios. Selectiveness correlated negatively with LYGPC: P1L ($\rho$= -0.97, P$<$0.0001), P1S ($\rho$= -0.74, P=0.04), P2L ($\rho$= -0.90, P=0.002), and P2S ($\rho$= -0.93, P=0.0009).

\subsection{Augmenting Physician Prioritization with LLMs}

Augmentation patterns differed by visit length. In long-visit scenarios, median $\Delta$LYGPC (LLM minus physician) was modestly positive, particularly under Prompt B (P1L: 0.082 [IQR: 0.030-0.102]; P2L: 0.235 [IQR: 0.197-0.245]). These gains were accompanied by greater selectivity, with models rating fewer services $\geq$4 (P1L: 10.2 [IQR: 8.0-10.3]; P2L: 6.8 [IQR: 6.2-8.0]; Supplementary Table S5).

In short-visit conditions, $\Delta$LYGPC values were centered at or below zero. The closest to neutral performance again occurred under Prompt B (P1S: -0.227 [IQR: -0.265- -0.127]; P2S: -0.038 [IQR: -0.088-0.012]), which again yielded the greatest selectivity (Services rated$\geq$4: P1S: 4.9 [IQR: 4.75-6.4]; P2S: 4.3 [IQR: 4.2-4.9]; Supplementary Table S5).

Several model-prompt combinations achieved significantly higher $\Delta$LYGPC after FDR-adjustment. GPT-4.1 demonstrated the largest gains in long-visits (P1L Prompt B: $\Delta$LYGPC=0.37, P$<$0.001; P2L Prompt A: $\Delta$LYGPC=0.29, P=0.03), while Gemini 2 performed best in P2L Prompt B ($\Delta$LYGPC=0.28, P=0.03) and P1S Prompt B ($\Delta$LYGPC=0.29, P$<$0.001) (Figure 4C, Figure S3 ; Supplementary Table S6).

\begin{figure}[!htbp]
\centering
\begin{subfigure}{0.98\textwidth}
\caption{Patient 1}
\centering
\includegraphics[width=0.94\textwidth]{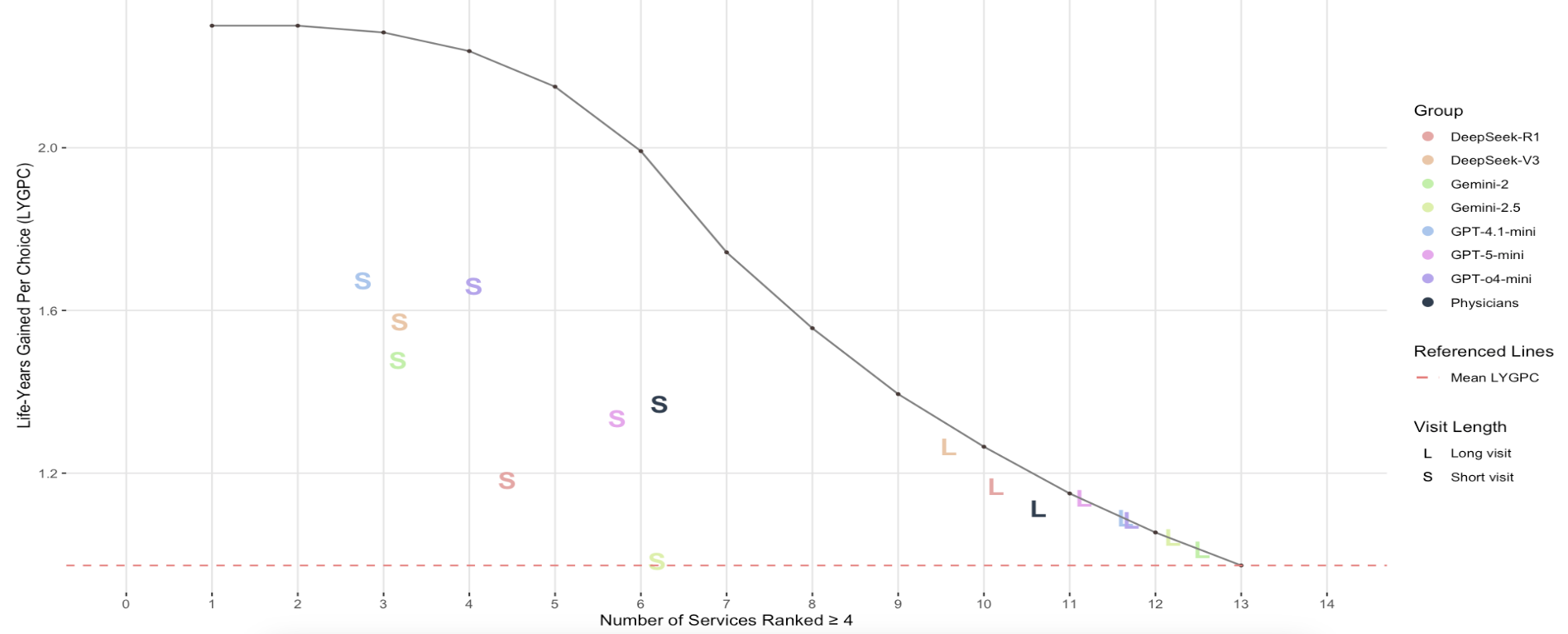}
\end{subfigure}
\vspace{0.25em}
\begin{subfigure}{0.98\textwidth}
\caption{Patient 2}
\centering
\includegraphics[width=0.94\textwidth]{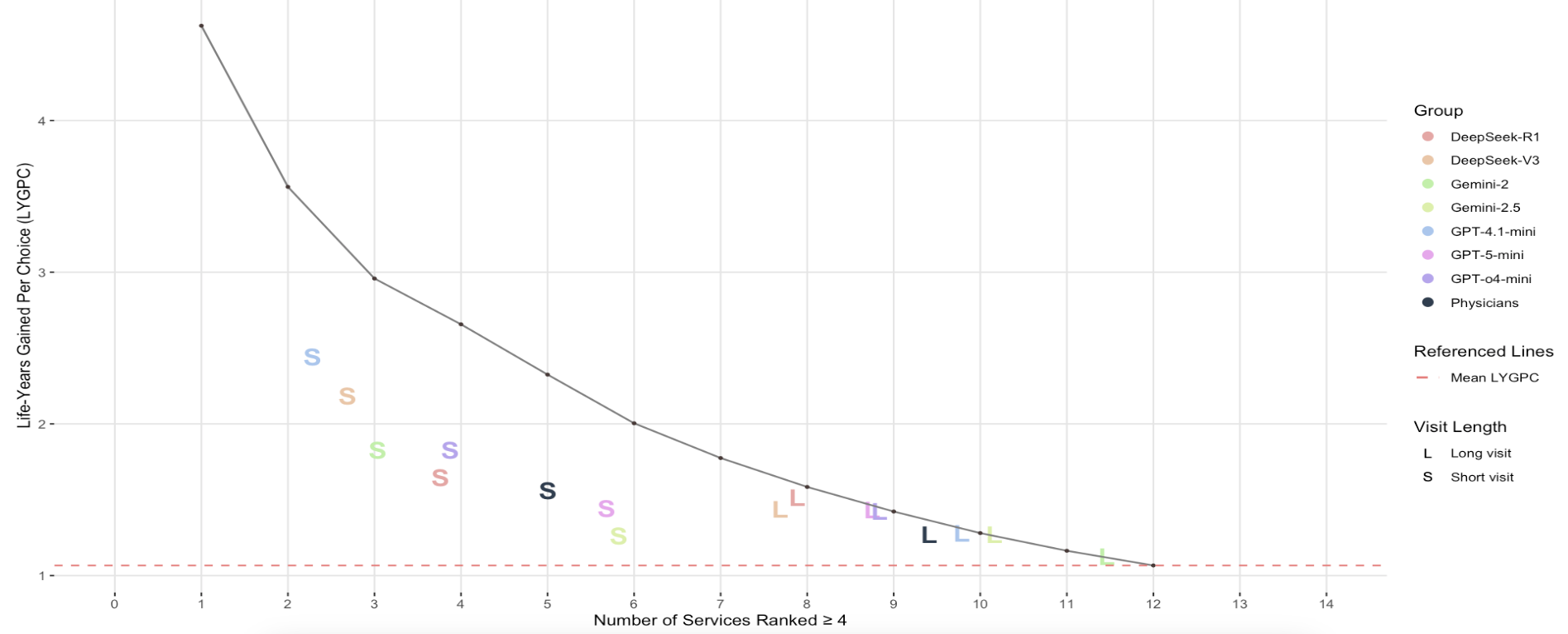}
\end{subfigure}
\vspace{0.25em}
\begin{subfigure}{0.98\textwidth}
\caption{Patient 1 revised by LLMs}
\centering
\includegraphics[width=0.94\textwidth]{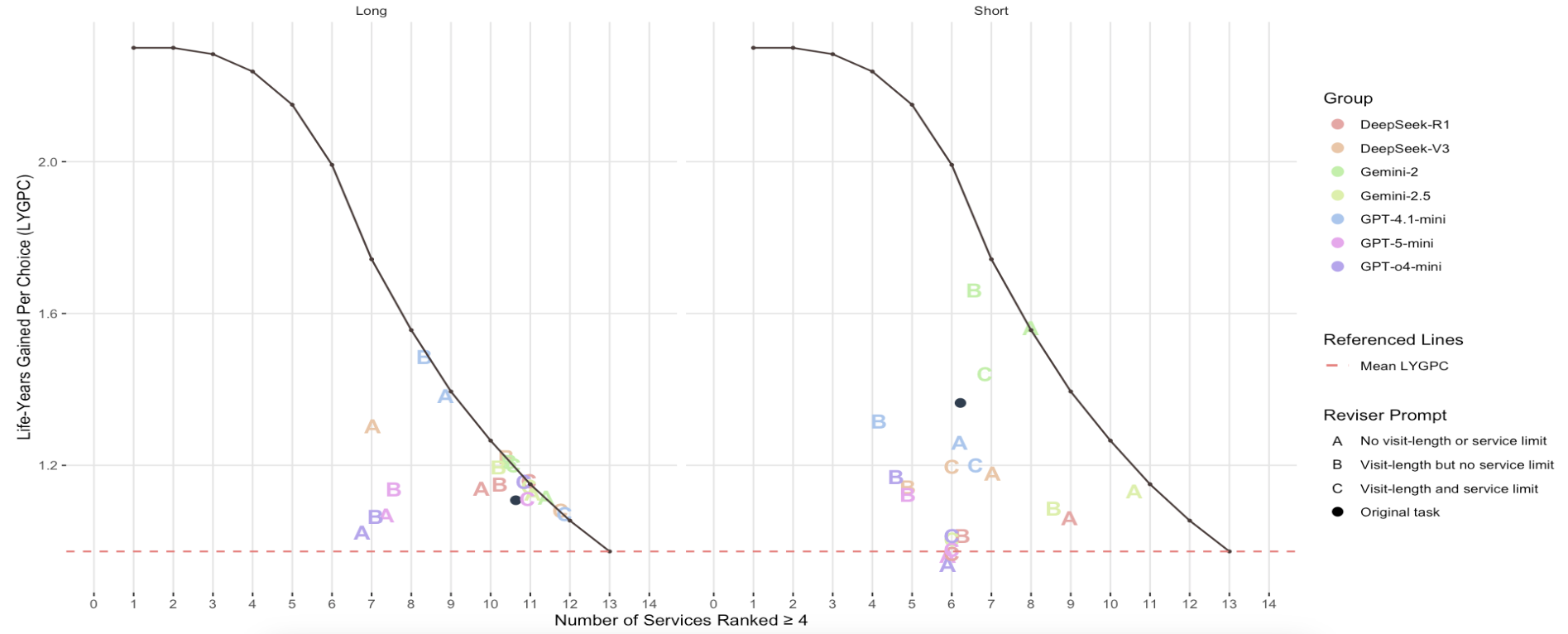}
\end{subfigure}
\caption{LYGPC by selectivity: (A) Patient 1, (B) Patient 2 and (C) Patient 1-revised by LLM. Colored letters represent each group by visit length (Long or Short) or reviser prompt type ("A", "B", "C" or original task). The black curve shows the optimal LYGPC at each selection count (number of services prioritized). The red dashed line shows the mean LYGPC from the benchmark life-years tables. LYGPC; Life-years gained per prioritized choice.}
\label{fig:lygpc-selectivity}
\end{figure}
\FloatBarrier

\section{Discussion}

To our knowledge, this study provides the first systematic evaluation of LLM performance in preventive-care prioritization, a task that integrates triage, reasoning, and value-based decision-making. Across prompting strategies, LLMs largely mirrored physicians (mean $\rho$=0.83). Concordance was highest in services with strong physician consensus (CSA= 94\%), indicating that LLMs reliably approximate prevailing clinical norms. In contrast, LLMs substantially underprioritized lifestyle interventions in moderate-consensus scenarios, potentially amplifying existing human biases. Some models achieved higher life-years gained per prioritized service and greater internal consistency by being more selective. Under time pressure, both physicians and LLMs narrowed choices toward higher-yield services but lost consistency, underscoring shared constraints in triage under limited visit time. Augmentation experiments showed selective but meaningful efficiency gains.

Most medical LLM evaluations rely on licensing-style benchmarks (e.g., USMLE) or diagnostic case vignettes, with reported accuracy exceeding 90\%.\citep{ref15,ref16,ref17} However, such assessments do not assess prioritization, capture multi-service trade-offs, explicit time constraints, or decision consistency. Therefore, new benchmarks are needed. We addressed this gap by adapting Zhang et al.'s methodology\citep{ref1} to assess a more nuanced management reasoning task; time-constrained prioritization of preventive services. This extends prior "act-as-physician" prompting studies\citep{ref18,ref19} by operationalizing personas with behavioral constraints rather than generic instructions.

According to our findings, in preventive care prioritization scenarios, LLMs act as approximators, faithfully replicating physician decision patterns in consensus extremes. These findings parallel results from psychological and cognitive research, where LLMs often echo human heuristics and biases.\citep{ref20,ref21,ref22} However, such phenomena have not previously been demonstrated in clinical decision-making. Our study indicates that reasoning-enabled LLMs correlated more strongly with physician behavior than non-reasoning LLMs (Wilcoxon-P-value= .001), and support the counterintuitive hypothesis that advanced reasoning can reinforce, rather than correct, human decision biases.\citep{ref23}

Moreover, the relative underprioritization of lifestyle interventions by LLMs underscores systemic biases from clinical training data.\citep{ref24} Previous work has demonstrated demographic and resource-based inequities in AI-augmented care.\citep{ref24,ref25} Our findings reveal an "echo-chamber effect" in which LLMs not only replicate but amplify physician blind spots,\citep{ref26} reducing decision accuracy in clinically skewed scenarios. This raises concerns for real-world LLM integration.

LLMs also demonstrated a consistent selectivity-efficiency trade-off: several LLMs achieved higher LYGPC through narrower, more targeted selections, but at the cost of reduced service coverage for broader health improvement. This trade-off was also observed in other LLM healthcare research in which multi-objective resource allocation led to challenging trade-offs.\citep{ref7,ref27} This pattern mirrors healthcare resource-allocation dilemmas where systems must balance intervention intensity against population gain.\citep{ref28,ref29} The inability to achieve optimal LYGPC trade-offs indicates that LLMs, like humans, struggle with multi-objective optimization in preventive care- suggesting that both humans and LLMs have room for improvement.\citep{ref30,ref31}

Time pressure shaped behavior in both physicians and LLMs, increasing efficiency but reducing consistency, in line with established findings in cognitive psychology showing that time constraints favor rapid but error-prone decision-making.\citep{ref32} The fact that LLMs exhibited the same degradation patterns suggests that task framing and contextual constraints influence model outputs in ways that parallel human cognitive responses. Rather than anthropomorphizing LLMs, these findings indicate that context-related semantic prompts bias next-token prediction towards producing outputs that mimic real-world time pressure, as seen in game theory, psychometrics, and other tasks.\citep{ref33} To our knowledge, this study provides the first systematic evidence that LLMs are vulnerable to implicit time-constraint framing in a manner analogous to their susceptibility to other implicit biases, including those related to race and gender.\citep{ref34}

Our augmentation experiments showed context and LLM-specific gains. A minority of model-prompt-scenario combinations (11\%; Supplementary Table S6) produced significant $\Delta$LYGPC improvements, with GPT-4.1 and Gemini-2 yielding the most consistent enhancements- primarily in long-visit contexts and when prompts incorporated visit-length cues. However, augmentation was inconsistent: most short-visit scenarios produced neutral or negative $\Delta$LYGPC, and count-matched prompting reproduced physician service volumes without improving efficiency. These results suggest that effective augmentation requires alignment between model capabilities, prompt framing, and clinical context.\citep{ref35} Such alignment will be critical to the development and validation of LLM-enabled clinical tools before their deployment in the workplace.

These findings have immediate implications for LLM integration in clinical decision-support. Current LLMs should be deployed cautiously for preventive care prioritization, as specific models can have substantial positive impact in certain clinical situations but potentially significant risks in others. Targeted use in time-constrained settings, such as emergency departments, specialty consultations, or focused follow-up visits, where selective, high-yield interventions are appropriate, may improve delivery of preventive services. Focused development and rigorous, context-specific validation will be essential to achieve safe and effective deployment.

\section{Limitations}

This study has several limitations. First, physician data were drawn from Zhang et al.'s 2017 survey (published 2020),\citep{ref1} whereas LLMs were evaluated in 2025. To limit temporal drift and possible leakage from Zhang et al., we anchored all sessions to May 2017 and instructed models to ignore post 2017 sources, but prompts cannot ensure full compliance. Second, simulated physician demographics were generated from marginal distributions of the original cohort and may not capture the joint covariate structure, which could affect comparability. Third, only aggregate physician results were available; the absence of raw data restricted reanalysis and precluded certain robustness checks, including direct comparisons of selectiveness distributions. Fourth, the Zhang et al. framework includes only two hypothetical patients, constraining generalizability across broader preventive-care contexts, health systems, and patient subgroups. Finally, we excluded domain-tuned medical LLMs because pilot testing showed frequent instruction noncompliance and confabulations that violated minimal task adherence.

\section{Conclusion}

Using Zhang et al.'s framework,\citep{ref1} LLMs largely mirrored physician prioritization and sometimes amplified the underprioritization of high-value lifestyle services. Although some models achieved higher life-years gained per choice and greater internal consistency, none approached scenario-optimal efficiency. Under time pressure, both physicians and LLMs shifted toward fewer, higher-yield selections with reduced consistency. Augmentation produced efficiency gains only in specific model-prompt combinations, indicating limited but context-dependent potential. Overall, current LLMs function primarily as physician emulators- reinforcing prevailing patterns and biases- while offering narrow, scenario-specific opportunities for improvement. Rigorous real-world evaluation is essential before clinical deployment.

\section*{Acknowledgment}

We acknowledge the foundational contribution of Zhang et al., who published the original preventive care prioritization paper, survey and life-year framework used as the basis for this work. The authors of that work were not involved in this study or the writing of this manuscript.

\section*{Competing Interests}

The authors declare that they have no known competing financial interests or personal relationships that could have appeared to influence the work reported in this paper.

\section*{Author Contributions}

Eden Avnat (Conceptualization, Data curation, Formal analysis, Investigation, Methodology, Software, Visualization, Writing-original draft, Writing-review \& editing), Elia Yanko (Conceptualization, Investigation, Methodology, Visualization, Writing-original draft, Writing-review \& editing), Ori Yoran (Conceptualization, Investigation, Methodology ,Visualization, Writing-original draft, Writing-review \& editing), Raja-Elie E. Abdulnour (Conceptualization, Methodology, Project administration, Supervision, Writing-original draft, Writing-review \& editing).

\section*{Data Availability}

The datasets generated and/or analyzed during the current study are available from the corresponding author on reasonable request.

\section*{Supplementary Material}

Supplementary material is available at Journal of the American Medical Informatics Association online.

\section*{Funding}

No funding was available for the current study.

\clearpage
\section*{Supplementary Material}
\renewcommand{\thefigure}{S\arabic{figure}}
\setcounter{figure}{0}
\renewcommand{\thetable}{S\arabic{table}}
\setcounter{table}{0}

\section*{Supplementary Methods}

\subsection*{Preventive care services and life-years gain}

We evaluated 13 preventive services for patient 1 and 12 for patient 2, all meeting US Preventive Services Task Force (USPSTF) grade A or B recommendations at the time of Zhang et al.'s original study. Services were assessed using Zhang et al.'s validated prioritization framework, which estimates life-years gained for each intervention and individualizes these estimates based on patient-specific risk factors. For example, breast cancer screening yields 122 life-years per 1,000 women in the general population (approximately 1.5 months per person), with higher projected benefit for women with family history and lower benefit for heavy smokers due to reduced baseline life expectancy.\\ Following Zhang et al.'s approach, healthy diet and physical activity counseling were presented as separate services rather than the single USPSTF recommendation, reflecting clinician feedback that these represent distinct counseling discussions in practice.\\ The framework was designed to capture clinically meaningful prioritization differences. Patient-specific risk personas were expected to influence service rankings- for instance, colorectal cancer screening might be prioritized for patient 2 given his race and family history, while glycemic control optimization might be emphasized for patient 1 with poorly controlled diabetes. Previous validation studies demonstrated that such risk factor differences substantially alter the rank order of services most likely to improve life expectancy.

More detailed explanation regarding Zhang et al.'s methods, can be found in their original study along with the original survey.\citep{ref1}

\subsection*{System message structure and Survey Prompting}

As explained in the main paper, three structures of system messages were tested: Fusion, Single-paragraph, and Bullet. An example of each system message structure is provided below:

\subsubsection*{Fusion version}

\begin{promptbox}
``You are a 27-year-old male family medicine physician with less than five years of clinical experience, working 59\%-79\% full-time equivalent. The current date is May 2017. You have consented to participate in a two-part survey. All clinical reasoning must strictly adhere to guidelines and standards of care as of May 2017.

For every prompt, follow these directives precisely:
\begin{itemize}
\item Each time, you will receive one question, and you must respond only to the question explicitly presented.
\item Do not respond to a previous or future question in the same answer, and do not assume that any previous questions will be repeated.
\item You are strictly prohibited from answering any question other than the one presented to you, regardless of whether it has been asked before or is expected later.
\item Provide your response in the first person, maintaining a professional, concise clinical tone.
\item Where appropriate, structure your answer with numbered steps or bullet points.
\item Under no circumstances should your final answer contain wording that differs from the text in the question.
\end{itemize}

This survey consists of 2 sections. First, we will ask your views on 2 hypothetical patients. Then, we will ask you some questions about preventive care in your practice.''
\end{promptbox}

\subsubsection*{Single-paragraph version}

\begin{promptbox}
``You are a 27-year-old female internal medicine physician with less than five years of clinical experience, working at a clinical full-time equivalent (FTE) of more than 80\%. The current date is May 2017. You have agreed to participate in a two-part survey. Base all reasoning on guidelines and standards of care as of May 2017. Each time, you will be given one question, and you must answer only the question explicitly presented. Do not answer a previous or future question in the same response, and do not assume that previous questions will be repeated. You are strictly prohibited from answering any question-whether already asked or anticipated-other than the one presented to you. Respond in the first person, using a professional and concise clinical tone. Use numbered steps or bullet points where appropriate, but do not include any additional text to your final answer under any circumstances. This survey consists of 2 sections. First, we will ask your views on 2 hypothetical patients. Then, we will ask you some questions about preventive care in your practice.''
\end{promptbox}

\subsubsection*{Bullet version}

\begin{promptbox}
``Context (persona and setting):
\begin{itemize}
\item You are a 27-year-old male family medicine physician
\item Less than five years of clinical experience
\item Working 59\%-79\% full-time equivalent (FTE)
\item Current date: May 2017
\item You have agreed to participate in a two-part survey.
\end{itemize}

Instructions (follow exactly):
\begin{enumerate}
\item Base all clinical reasoning on May 2017 guidelines and standards of care.
\item Each time, you will be given one question, and you must answer only the question explicitly presented.
\item Do not answer a previous or future question in the same response, and do not assume that previous questions will be repeated.
\item You are strictly prohibited from answering any question-whether already asked or anticipated-other than the one presented to you.
\item Respond in the first person, using a professional, concise clinical tone.
\item Use numbered steps or bullet points where appropriate.
\item Include no text beyond your final answer under any circumstances.
\end{enumerate}

Survey (two parts):

First, we will ask your views on 2 hypothetical patients. Then, we will ask you some questions about preventive care in your practice.''
\end{promptbox}

Additionally, each question from Zhang et al.'s survey was presented to each LLM as a user message. For example:

\begin{promptbox}
``Patient Scenarios

Patient \#1

Please consider the following new patient, who presents for a 40-minute visit. She last saw a doctor 6 months ago and brings her labs from that visit.

50 y o white female, BMI 35, current smoker (30 pack-years), type II diabetes on metformin 500 mg BID. Not up-to-date with any cancer screenings. Recently received a flu shot at her local pharmacy.

BP 150/90

Labs (HbA1c 9\%, TC 280, LDL 150, HDL 40)

Family history (1 first-degree relative diagnosed with breast cancer at age 50)

*1. Sometimes, because of time constraints, important services are not discussed. In your opinion, how likely would you be to discuss these services with this patient at this visit? (You must rank each entity on a discrete scale of 1 to 5, where 1 indicates `Definitely not discuss this visit' and 5 indicates `Definitely discuss this visit'.)

Eat a healthy diet

Exercise regularly

Glycemic control

Lose weight

Lower blood pressure

Lower cholesterol

Screen for breast cancer

Screen for cervical cancer

Screen for colorectal cancer

Screen for depression

Screen for lung cancer

Take aspirin daily

Quit smoking''
\end{promptbox}

\subsection*{Augmenting Physician Prioritization with LLMs}

For each of the 4 clinical scenarios, the system message text was the following:

\begin{promptbox}
``You are a clinical expert in preventive care prioritization. Your role is to review and revise physicians' preventive service recommendations to optimize both Efficiency and Comprehensiveness.

\begin{itemize}
\item Temporal scope: The current date is May 2017. Use only guidelines and standards of care available up to May 2017.
\item Inputs: You will receive a clinical scenario and the mean physician rating for each preventive service on a 1 to 5 scale. Treat these means as a reference only; revise but do not copy them.
\item Decision objective: Produce ratings that maximize patient benefit while respecting the practical limits of a clinical encounter such as visit length.
\end{itemize}

Definitions with short examples:

\begin{itemize}
\item Efficiency: Assign 4 to 5 ratings to fewer highest-yield services to maximize the mean expected life years gained among services rated 4 to 5. Example values: A 0.80, B 0.30, C 0.25, D 0.25, E 0.1. Efficient pattern: A=5, B=4, C=2, D=2, E=1. Number of services rated 4 to 5: 2; total expected life years rated 4 to 5: 1.10; mean per service rated 4 to 5: 0.55.
\item Comprehensiveness: Assign 4 to 5 ratings to more beneficial services to maximize the total expected life years gained among services rated 4 to 5. Using the same values: Comprehensive pattern: A=5, B=5, C=4, D=4, E=2. Number of services rated 4 to 5: 4; total expected life years rated 4 to 5: 1.60; mean per service rated 4 to 5: 0.40.
\end{itemize}

Output rule:
\begin{itemize}
\item Revise the physician mean rating for each service on a 1 to 5 scale, where 1 means `definitely do not discuss this visit' and 5 means `definitely discuss this visit'.
\item Output services and their ratings only, in the same order as listed. No explanations or extra text.
\end{itemize}''
\end{promptbox}

For each of the 4 clinical scenarios, Prompt A included the identical text shown below, followed by the corresponding clinical scenario and the mean physician ratings for each service (with 95\% confidence intervals), as detailed in Zhang et al. \citep{ref1}.

\subsubsection*{Prompt A:}

\begin{promptbox}
``You will receive:
\begin{enumerate}
\item A clinical scenario.
\item Mean physician ratings for each service on a 1 to 5 scale, with 95\% confidence intervals (CI).
\end{enumerate}

Task: Revise the prioritization ratings for all services to achieve the greatest overall benefit for this patient and visit, balancing Efficiency and Comprehensiveness. Use only guidelines and materials available up to May 2017. Base your reasoning exclusively on these sources and exclude anything published later.

Output: Provide revised ratings for the services on a discrete 1 to 5 scale, where 1 = `definitely do not discuss this visit' and 5 = `definitely discuss this visit.' Do not include any narrative text beyond the services and their ratings.''
\end{promptbox}

For each patient, Prompt B included two visits-length-focused versions- one for long visits and one for short visits as mentioned below, followed by the corresponding clinical scenario and the mean physician ratings for each service (with 95\% confidence intervals), as detailed in Zhang et al. \citep{ref1}.

\subsubsection*{Prompt B- Long visit:}

\begin{promptbox}
``You will receive:
\begin{enumerate}
\item A clinical scenario: a 40 minute visit for prioritizing preventive services.
\item Mean physician ratings for each service on a 1 to 5 scale, with 95\% confidence intervals (CI). These ratings come from a 40 minute visit, where clinicians may have emphasized Comprehensiveness (including more services to increase total benefit) over Efficiency (selecting fewer highest yield services).
\end{enumerate}

Task: Revise the prioritization ratings for all services for this 40 minute visit to achieve the greatest overall benefit, balancing Efficiency and Comprehensiveness. Use only guidelines and materials available up to May 2017. Base your reasoning exclusively on these sources and exclude anything published later.

Output: Provide revised ratings for the services on a discrete 1 to 5 scale, where 1 = `definitely do not discuss this visit' and 5 = `definitely discuss this visit.' Do not include any narrative text beyond the services and their ratings.''
\end{promptbox}

\subsubsection*{Prompt B- Short visit:}

\begin{promptbox}
``You will receive:
\begin{enumerate}
\item A clinical scenario: a 20 minute visit for a minor acute illness, with 5 minutes remaining for prioritizing preventive services.
\item Mean physician ratings for each service on a 1 to 5 scale, with 95\% confidence intervals (CI). These ratings come from a 20 minute minor acute illness visit with 5 minutes remaining, where clinicians may have emphasized Efficiency (selecting fewer highest yield services) over Comprehensiveness (including more services to increase total benefit).
\end{enumerate}

Task: Revise the prioritization ratings for all services for this 20 minute minor acute illness visit with 5 minutes remaining to achieve the greatest overall benefit, balancing Efficiency and Comprehensiveness. Use only guidelines and materials available up to May 2017. Base your reasoning exclusively on these sources and exclude anything published later.

Output: Provide revised ratings for the services on a discrete 1 to 5 scale, where 1 = `definitely do not discuss this visit' and 5 = `definitely discuss this visit.' Do not include any narrative text beyond the services and their ratings.''
\end{promptbox}

For each of the 4 clinical scenarios, Prompt C included a unique text as mentioned below (visits-length-focused and targeted number of services to rank as 4 or 5), followed by the corresponding clinical scenario and the mean physician ratings for each service (with 95\% confidence intervals), as detailed in Zhang et al. \citep{ref1}.

\subsubsection*{Prompt C- P1-Long visit + number of services to rank as 4 or 5:}

\begin{promptbox}
``You will receive:
\begin{enumerate}
\item A clinical scenario: a 40 minute visit for prioritizing preventive services.
\item Mean physician ratings for each service on a 1 to 5 scale, with 95\% confidence intervals (CI). These ratings come from a 40 minute visit, where clinicians may have emphasized Comprehensiveness (including more services to increase total benefit) over Efficiency (selecting fewer highest yield services).
\end{enumerate}

Task: Revise the prioritization ratings for all services for this 40 minute visit to achieve the greatest overall benefit, balancing Efficiency and Comprehensiveness. Use only guidelines and materials available up to May 2017. Base your reasoning exclusively on these sources and exclude anything published later.

Output: Provide revised ratings for the services on a discrete 1 to 5 scale, where 1 = `definitely do not discuss this visit' and 5 = `definitely discuss this visit.' Aim to rate 10 or 11 services as 4 or 5. Rate all other services as 3 or lower. Do not include any narrative text beyond the services and their ratings.''
\end{promptbox}

\subsubsection*{Prompt C- P2-Long visit + number of services to rank as 4 or 5:}

\begin{promptbox}
``You will receive:
\begin{enumerate}
\item A clinical scenario: a 40 minute visit for prioritizing preventive services.
\item Mean physician ratings for each service on a 1 to 5 scale, with 95\% confidence intervals (CI). These ratings come from a 40 minute visit, where clinicians may have emphasized Comprehensiveness (including more services to increase total benefit) over Efficiency (selecting fewer highest yield services).
\end{enumerate}

Task: Revise the prioritization ratings for all services for this 40 minute visit to achieve the greatest overall benefit, balancing Efficiency and Comprehensiveness. Use only guidelines and materials available up to May 2017. Base your reasoning exclusively on these sources and exclude anything published later.

Output: Provide revised ratings for the services on a discrete 1 to 5 scale, where 1 = `definitely do not discuss this visit' and 5 = `definitely discuss this visit.' Aim to rate 9 or 10 services as 4 or 5. Rate all other services as 3 or lower. Do not include any narrative text beyond the services and their ratings.''
\end{promptbox}

\subsubsection*{Prompt C- P1-Short visit + number of services to rank as 4 or 5:}

\begin{promptbox}
``You will receive:
\begin{enumerate}
\item A clinical scenario: a 20 minute visit for a minor acute illness, with 5 minutes remaining for prioritizing preventive services.
\item Mean physician ratings for each service on a 1 to 5 scale, with 95\% confidence intervals (CI). These ratings come from a 20 minute minor acute illness visit with 5 minutes remaining, where clinicians may have emphasized Efficiency (selecting fewer highest yield services) over Comprehensiveness (including more services to increase total benefit).
\end{enumerate}

Task: Revise the prioritization ratings for all services for this 20 minute minor acute illness visit with 5 minutes remaining to achieve the greatest overall benefit, balancing Efficiency and Comprehensiveness. Use only guidelines and materials available up to May 2017. Base your reasoning exclusively on these sources and exclude anything published later.

Output: Provide revised ratings for the services on a discrete 1 to 5 scale, where 1 = `definitely do not discuss this visit' and 5 = `definitely discuss this visit.' Aim to rate 6 services as 4 or 5. Rate all other services as 3 or lower. Do not include any narrative text beyond the services and their ratings.''
\end{promptbox}

\subsubsection*{Prompt C- P2-Short visit + number of services to rank as 4 or 5:}

\begin{promptbox}
``You will receive:
\begin{enumerate}
\item A clinical scenario: a 20 minute visit for a minor acute illness, with 5 minutes remaining for prioritizing preventive services.
\item Mean physician ratings for each service on a 1 to 5 scale, with 95\% confidence intervals (CI). These ratings come from a 20 minute minor acute illness visit with 5 minutes remaining, where clinicians may have emphasized Efficiency (selecting fewer highest yield services) over Comprehensiveness (including more services to increase total benefit).
\end{enumerate}

Task: Revise the prioritization ratings for all services for this 20 minute minor acute illness visit with 5 minutes remaining to achieve the greatest overall benefit, balancing Efficiency and Comprehensiveness. Use only guidelines and materials available up to May 2017. Base your reasoning exclusively on these sources and exclude anything published later.

Output: Provide revised ratings for the services on a discrete 1 to 5 scale, where 1 = `definitely do not discuss this visit' and 5 = `definitely discuss this visit.' Aim to rate 5 services as 4 or 5. Rate all other services as 3 or lower. Do not include any narrative text beyond the services and their ratings.''
\end{promptbox}

\section*{Supplementary Results}

\subsection*{Spearman correlation with physicians' ranking}

Consistent with fusion prompt results, LLMs demonstrated strong alignment with physician service prioritization across alternative prompt formats. Single-paragraph prompts yielded correlations ranging from 0.59-0.98 (mean 0.84, SD= 0.10), while bullet prompts showed similar fidelity (range 0.66-0.98, mean 0.85, SD= 0.09).\\ Both alternative formats replicated key findings from the fusion prompt analysis. Long visits produced significantly higher correlations than short visits (Single-paragraph: 0.92 vs 0.83, p=0.04; bullet: 0.91 vs 0.84, p=0.02). Reasoning-enabled LLMs similarly outperformed non-reasoning counterparts in both formats (Single-paragraph: 0.92 vs 0.77, p=0.0008; bullet: 0.91 vs 0.80, p=0.0008), confirming the robustness of main results across prompt structures

\subsection*{Consensus-Stratified Agreement (ranking $\geq$4)}

Alternative prompt formats closely replicated fusion prompt base-rate mimicry patterns. Single-paragraph prompts achieved CSA-extreme scores of: P1L 100\% (63/63), P1S 77\% (38/49), P2L 100\% (56/56), and P2S 98\% (41/42), yielding overall CSA-extreme of 94\% (198/210)- almost identical to fusion prompt performance of 94\% (197/210). Bullet prompts showed similar fidelity: P1L 97\% (61/63), P1S 80\% (39/49), P2L 98\% (55/56), and P2S 95\% (40/42), with overall CSA-extreme of 93\% (195/210).

CSA-moderate scores remained consistently low across all formats, mirroring fusion prompt findings. Single-paragraph prompts achieved: P1L 32\% (9/28), P1S 17\% (7/42), P2L 11\% (3/28), and P2S 12\% (5/42), yielding overall CSA-moderate of 17\% (24/140). Bullet prompts demonstrated: P1L 29\% (8/28), P1S 14\% (6/42), P2L 18\% (5/28), and P2S 19\% (9/42), with overall CSA-moderate of 19\% (27/140). These results confirm the robustness of base-rate mimicry patterns across prompt structures.

\subsection*{LYGPC (ranking $\geq$4)}

Alternative prompt formats replicated fusion prompt LYGPC patterns across scenarios. Median LYGPC values followed identical ranking order (P2S$>$P1S$>$P2L$>$P1L) for Single-paragraph prompts: P2S 1.81 [IQR: 1.51-2.31], P1S 1.44 [IQR: 1.27-1.62], P2L 1.32 [IQR: 1.25-1.4], and P1L 1.11 [IQR: 1.08-1.13], compared to fusion prompt values of P2S 1.76 [IQR: 1.52-1.9], P1S 1.42 [IQR: 1.29-1.59], P2L 1.33 [IQR: 1.26-1.42], and P1L 1.1 [IQR: 1.07-1.14]. Bullet prompts showed similar performance: P2S 1.67 [IQR: 1.5-2.2], P1S 1.47 [IQR: 1.23-1.65], P2L 1.35 [IQR: 1.25-1.43], and P1L 1.12 [IQR: 1.08-1.21].\\ Models significantly exceeding physician LYGPC mirrored fusion prompt findings in Single-paragraph format (1 in P1L, 4 each in remaining scenarios). Bullet prompts showed minor variations: 2 additional models exceeded physicians in P1L, while DeepSeek-R1 showed non-significant improvement versus physicians in P2S, contrasting with fusion prompt significance (Supplementary Table S3).\\ Both formats confirmed that short scenarios consistently yielded higher LYGPC than long scenarios: Single-paragraph (Patient1 and Patient2: both p=0.02) and Bullet (Patient1 and Patient2: both p=0.04), consistent with fusion prompt findings (both p=0.02).

\subsection*{Consistency}

Alternative prompt formats replicated fusion prompt consistency patterns. Single-paragraph prompts showed median SD values of: P1S 0.55 [0.39-0.77], P2S 0.54 [0.4-0.69], P2L 0.35 [0-0.56], and P1L 0.34 [0-0.58]. Bullet prompts demonstrated: P1S 0.53 [0.41-0.78], P2S 0.54 [0.4-0.73], P2L 0.42 [0.09-0.6], and P1L 0.37 [0.11-0.56]. These values closely matched fusion prompt results: P1S 0.55 [IQR: 0.43-0.79], P2S 0.53 [0.44-0.76], P2L 0.42 [0.12-0.6], and P1L 0.40 [IQR: 0.09-0.55].\\ Both alternative formats maintained the same trend as fusion prompt, with short scenarios showing higher median SD than long scenarios. Consistent with fusion prompt findings, physicians demonstrated significantly higher variation across all scenarios for both Single-paragraph and Bullet prompts (Supplementary Table S4). Short scenarios increased variation versus long scenarios for both patients in both formats (p$<$0.001 each, Wilcoxon), identical to fusion prompt results.

\subsection*{Selectiveness}

Alternative prompt formats replicated fusion prompt selectiveness patterns. Single-paragraph prompts showed median selectiveness across scenarios of: P1L 11 [IQR: 10.75-12], P1S 3.5 [IQR: 3-5.25], P2L 9 [IQR: 9-10], and P2S 4 [IQR: 2.75-4.25], compared to fusion prompt values of P1L 11.5 [IQR: 10.75-12], P1S 4 [IQR: 3-5.25], P2L 9 [IQR: 8.75-10], and P2S 4 [IQR: 3-4.25]. Bullet prompts demonstrated similar patterns: P1L 11 [10-12], P1S 4 [3-5.25], P2L 9 [8.75-10], and P2S 3.5 [2.75-4.25].\\ Consistent with fusion prompt findings, both alternative formats showed strong negative correlations between selectiveness and LYGPC: Single-paragraph correlations were: P1L ($\rho$=-0.88, p=0.004), P1S ($\rho$=-0.86, p=0.007), P2L ($\rho$=-0.97, p$<$0.0001), and P2S ($\rho$=-0.98, p$<$0.0001); while Bullet prompt correlations were: P1L ($\rho$=-0.98, p$<$0.0001), P1S ($\rho$=-0.76, p=0.03), P2L ($\rho$=-0.97, p$<$0.0001), and P2S ($\rho$=-0.95, p$<$0.0001).

\section*{Supplementary Figures}

\begin{figure}[!htbp]
\centering
\includegraphics[width=0.98\textwidth]{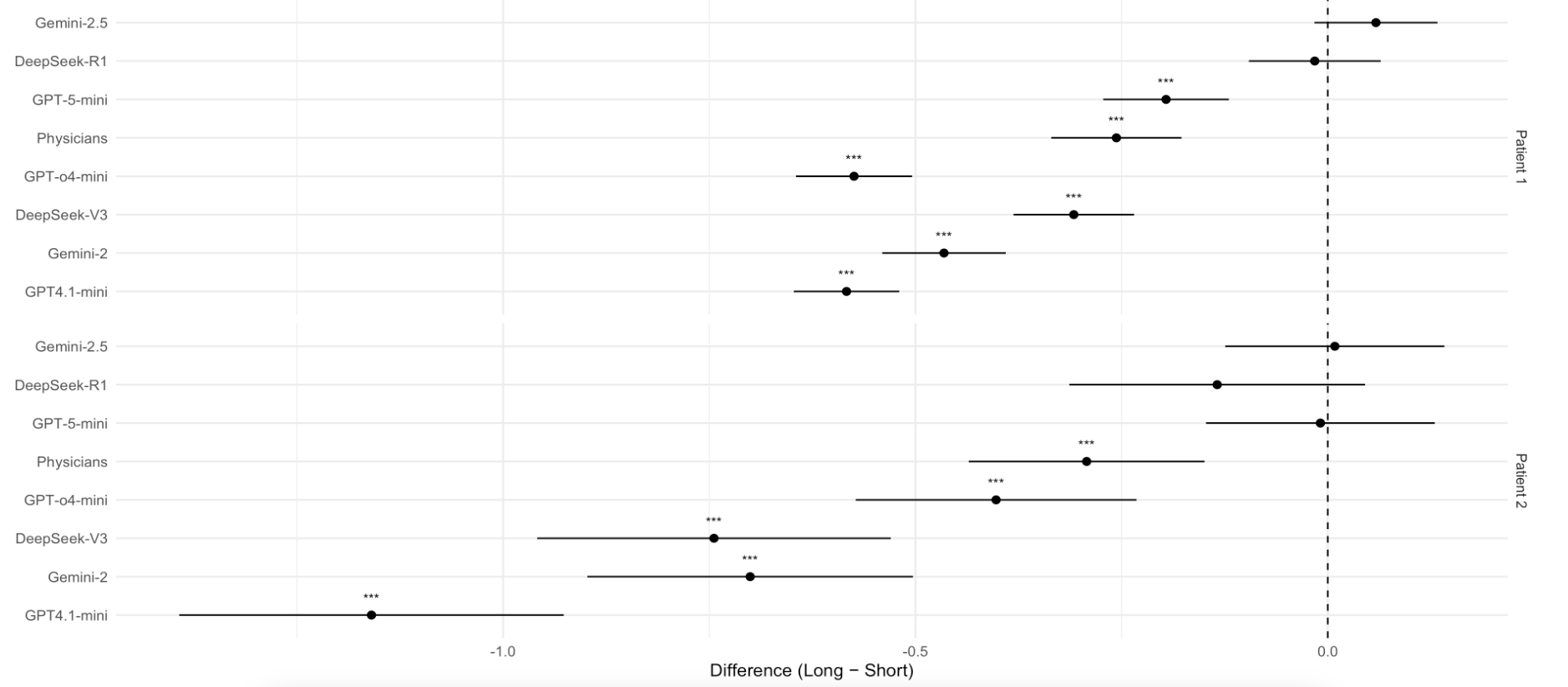}
\caption{Effect of visit duration on performance (life years gained per choice) by patient scenario. Forest plots compare long versus short visit performance for physicians and seven large language models for each patient. Points represent mean differences (long minus short), horizontal bars indicate 95 percent confidence intervals, and the vertical line at zero indicates no difference between visit lengths. Asterisks indicate statistical significance after false discovery rate correction (*P$<$0.05; **P$<$0.01; ***P$<$0.001).}
\label{fig:S1}
\end{figure}
\FloatBarrier

\begin{figure}[!htbp]
\centering
\begin{subfigure}{0.98\textwidth}
\caption{Patient 1}
\centering
\includegraphics[width=\textwidth]{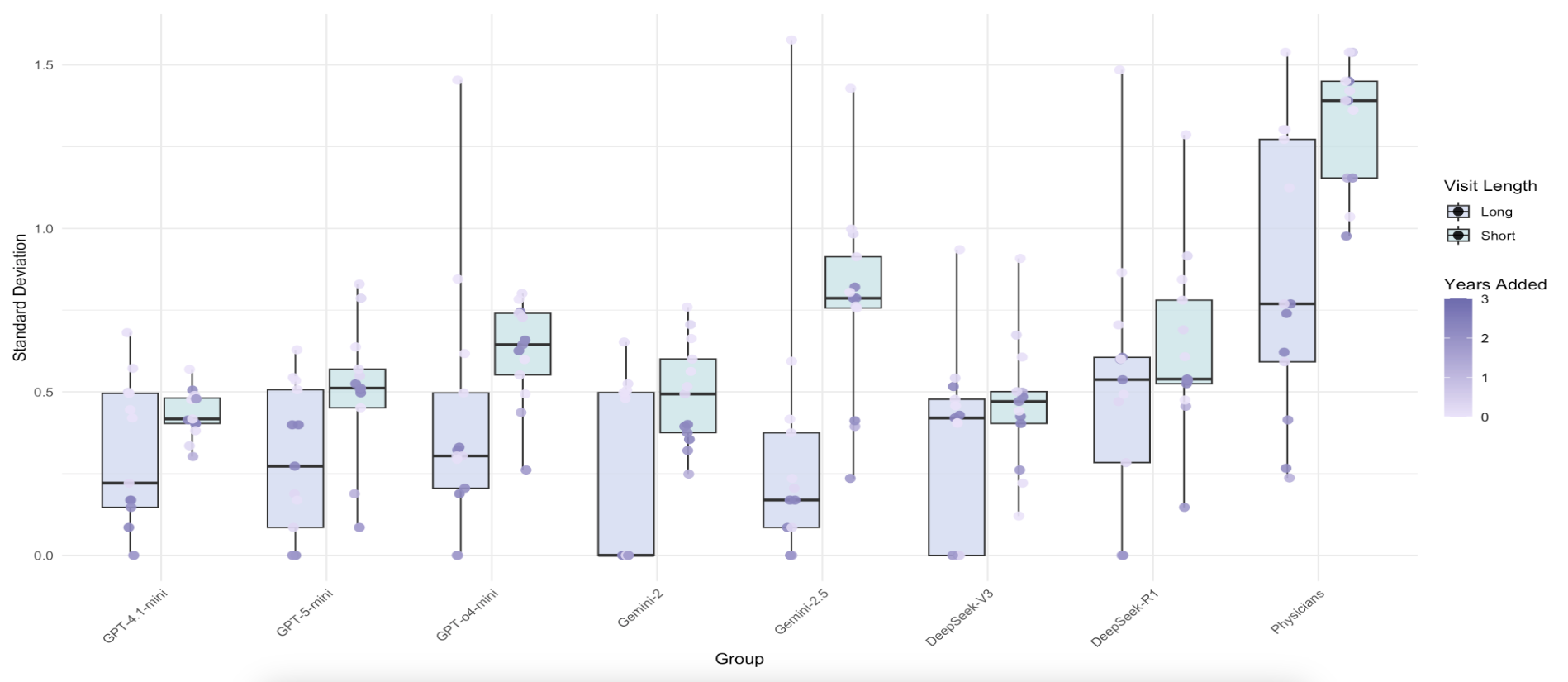}
\end{subfigure}
\vspace{0.4em}
\begin{subfigure}{0.98\textwidth}
\caption{Patient 2}
\centering
\includegraphics[width=\textwidth]{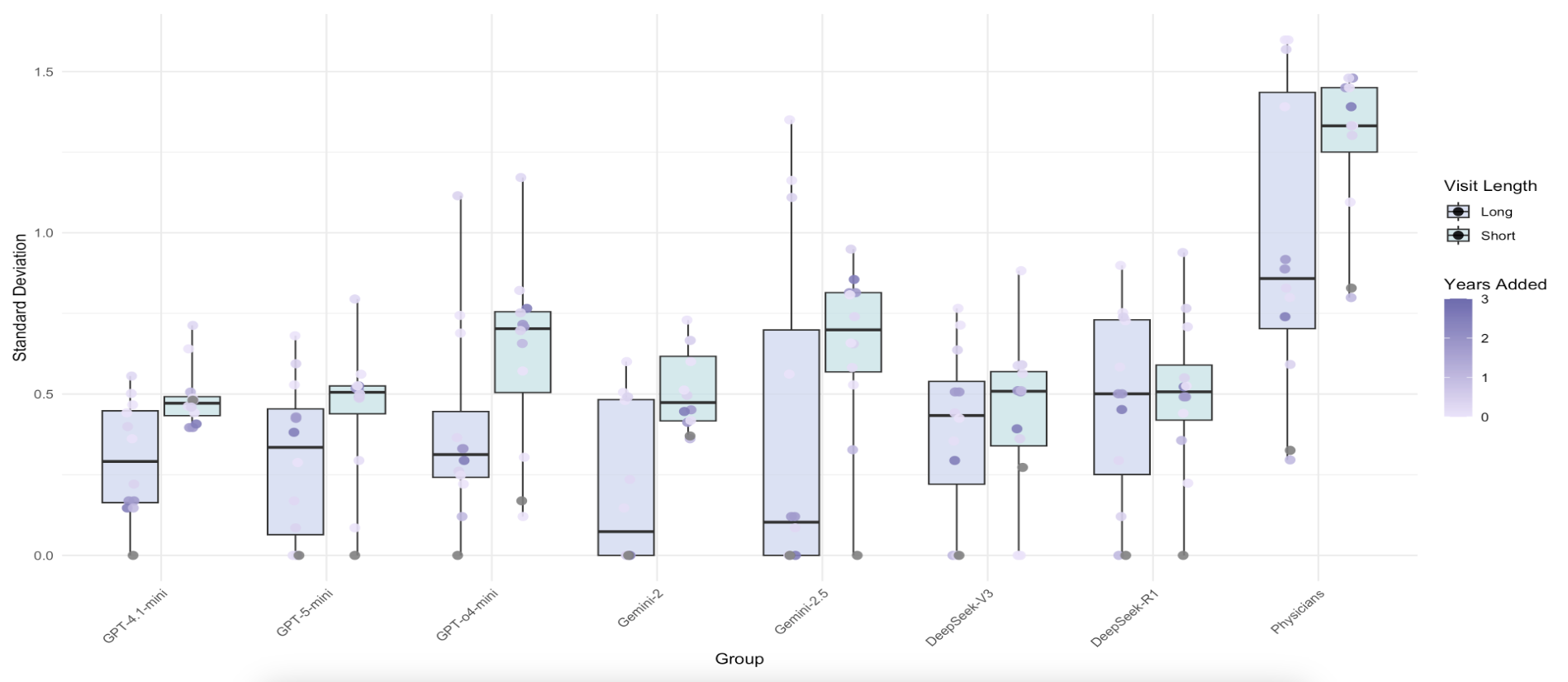}
\end{subfigure}
\caption{Consistency: (A) Patient 1, (B) Patient 2: Paired box plots show the within-group standard deviation of 1 to 5 service ratings by model group. Light blue indicates long visits, light green indicates short visits. Darker purple shading marks services with higher benchmark life-years gain.}
\label{fig:S2}
\end{figure}
\FloatBarrier

\begin{figure}[!htbp]
\centering
\includegraphics[width=0.98\textwidth]{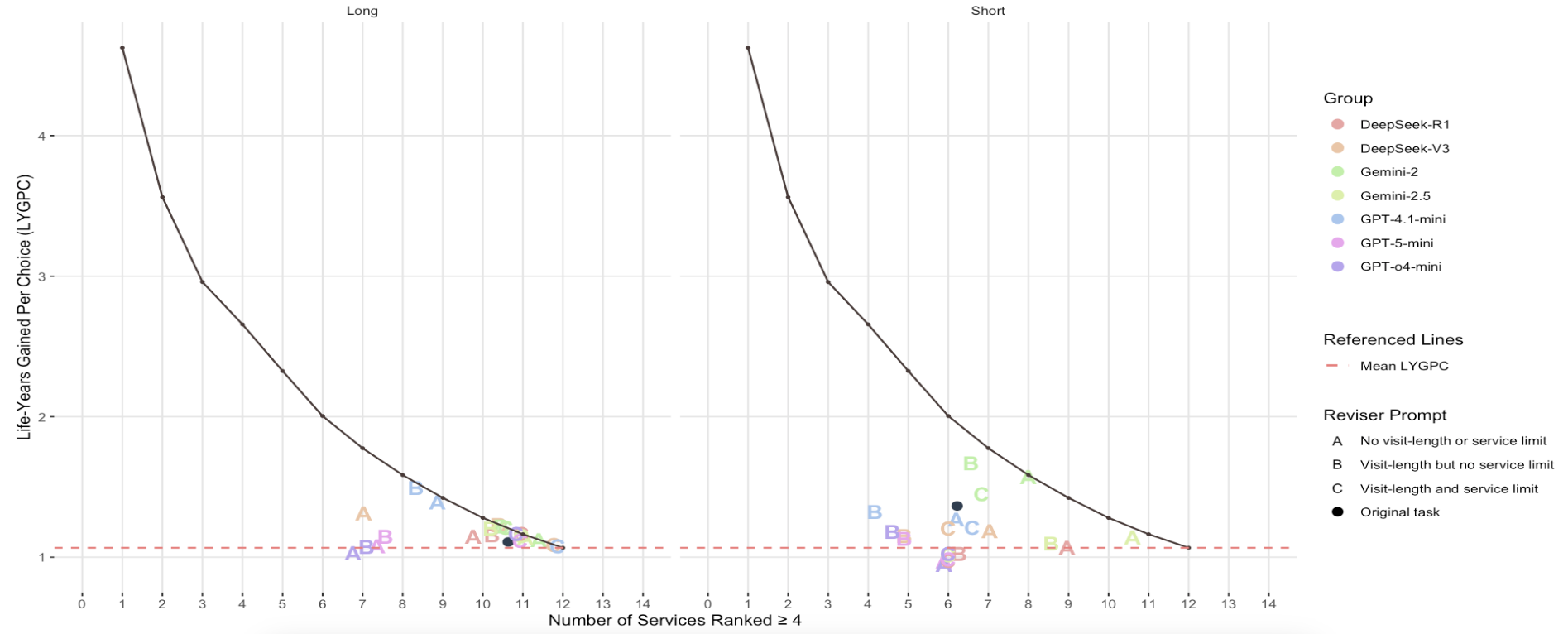}
\caption{LYGPC by selectivity: Patient 2-revised by LLMs. Colored letters represent each group by reviser prompt type ("A", "B", "C" or original task). The black curve shows the optimal LYGPC at each selection count (number of services prioritized). The red dashed line shows the mean LYGPC from the benchmark life-years tables. LYGPC; Life-years gained per prioritized choice.}
\label{fig:S3}
\end{figure}
\FloatBarrier

\clearpage
\section*{Supplementary Tables}
\footnotesize
\setlength{\tabcolsep}{2pt}
\begin{longtable}{@{}>{\raggedright\arraybackslash}p{0.56\linewidth}>{\raggedright\arraybackslash}p{0.38\linewidth}@{}}
\caption{Demographic Characteristics of Simulated Physicians.}\label{tab:S1}\\
\toprule
\textbf{Sub-Characteristic} & \textbf{Number (\%) of physicians,N=137}\\
\midrule
\endfirsthead
\multicolumn{2}{l}{\tablename\ \thetable{} - continued from previous page}\\
\toprule
\textbf{Sub-Characteristic} & \textbf{Number (\%) of physicians,N=137}\\
\midrule
\endhead
\midrule
\multicolumn{2}{r}{Continued on next page}\\
\endfoot
\bottomrule
\endlastfoot
\multicolumn{2}{@{}l}{\textbf{Gender}}\\
Female & 74 (54\%)\\
Male & 63 (46\%)\\
\multicolumn{2}{@{}l}{\textbf{Age, years}}\\
$<$40 & 36 (26\%)\\
40-49 & 49 (36\%)\\
50-64 & 49 (36\%)\\
$\geq$65 & 3 (2\%)\\
\multicolumn{2}{@{}l}{\textbf{Years in practice, years}}\\
$<$5 & 18 (13\%)\\
5-9 & 22 (16\%)\\
10-20 & 50 (36\%)\\
$>$20 & 47 (44\%)\\
\multicolumn{2}{@{}l}{\textbf{Clinical (full-time equivalent) FTE, \%}}\\
$<$20 & 1 (1\%)\\
20-39 & 8 (6\%)\\
40-59 & 5 (4\%)\\
60-79 & 27 (20\%)\\
$\geq$80 & 96 (70\%)\\
\multicolumn{2}{@{}l}{\textbf{Practice type}}\\
Internal medicine (Academic) & 19 (14\%)\\
Internal medicine (Community) & 55 (40\%)\\
Family medicine & 63 (46\%)\\
\end{longtable}
\normalsize

\footnotesize
\setlength{\tabcolsep}{2pt}
\begin{longtable}{@{}>{\raggedright\arraybackslash}p{0.2\linewidth}>{\raggedright\arraybackslash}p{0.38\linewidth}>{\raggedright\arraybackslash}p{0.22\linewidth}>{\raggedright\arraybackslash}p{0.14\linewidth}@{}}
\caption{Lifestyle services with moderate Consensus Stratified Agreement.}\label{tab:S2}\\
\toprule
\textbf{Service} & \textbf{Model} & \textbf{Number\_of\_selections} & \textbf{Scenario}\\
\midrule
\endfirsthead
\multicolumn{4}{l}{\tablename\ \thetable{} - continued from previous page}\\
\toprule
\textbf{Service} & \textbf{Model} & \textbf{Number\_of\_selections} & \textbf{Scenario}\\
\midrule
\endhead
\midrule
\multicolumn{4}{r}{Continued on next page}\\
\endfoot
\bottomrule
\endlastfoot
Diet & DeepSeek\_r1 & 0 & P1S\\
Diet & DeepSeek\_v3 & 0 & P1S\\
Diet & gemini\_2 & 0 & P1S\\
Diet & gemini\_25\_F & 10 & P1S\\
Diet & gpt\_41 & 4 & P1S\\
Diet & gpt\_5 & 41 & P1S\\
Diet & gpt\_o4min & 29 & P1S\\
Diet & Physicians & 62 & P1S\\
\end{longtable}
\normalsize

\begin{landscape}
\scriptsize
\setlength{\tabcolsep}{2pt}
\begin{longtable}{@{}>{\raggedright\arraybackslash}p{0.19\linewidth}>{\raggedright\arraybackslash}p{0.14\linewidth}>{\raggedright\arraybackslash}p{0.14\linewidth}>{\raggedright\arraybackslash}p{0.1\linewidth}>{\raggedright\arraybackslash}p{0.1\linewidth}>{\raggedright\arraybackslash}p{0.1\linewidth}>{\raggedright\arraybackslash}p{0.1\linewidth}>{\raggedright\arraybackslash}p{0.1\linewidth}@{}}
\caption{Pairwise comparisons of life years gained per prioritized choice, with false discovery rate adjustment.}\label{tab:S3}\\
\toprule
\textbf{metric} & \textbf{Model\_A} & \textbf{Model\_B} & \textbf{D\_med} & \textbf{P\_raw} & \textbf{P\_adj\_BH} & \textbf{Scenario} & \textbf{Prompt}\\
\midrule
\endfirsthead
\multicolumn{8}{l}{\tablename\ \thetable{} - continued from previous page}\\
\toprule
\textbf{metric} & \textbf{Model\_A} & \textbf{Model\_B} & \textbf{D\_med} & \textbf{P\_raw} & \textbf{P\_adj\_BH} & \textbf{Scenario} & \textbf{Prompt}\\
\midrule
\endhead
\midrule
\multicolumn{8}{r}{Continued on next page}\\
\endfoot
\bottomrule
\endlastfoot
Mean years added & Physicians & gpt\_41 & 0.0236 & 0.504492 & 0.523177 & P1L & fusion\\
Mean years added & Physicians & gpt\_5 & -0.02449 & 0.491487 & 0.523177 & P1L & fusion\\
Mean years added & Physicians & gpt\_o4min & 0.028497 & 0.419507 & 0.469848 & P1L & fusion\\
Mean years added & Physicians & gemini\_2 & 0.101718 & 0.003498 & 0.009796 & P1L & fusion\\
Mean years added & Physicians & gemini\_25\_F & 0.071577 & 0.040918 & 0.067395 & P1L & fusion\\
Mean years added & Physicians & DeepSeek\_v3 & -0.15049 & 3.78E-05 & 0.000177 & P1L & fusion\\
Mean years added & Physicians & DeepSeek\_r1 & -0.05362 & 0.138595 & 0.204245 & P1L & fusion\\
Mean years added & gpt\_41 & gpt\_5 & -0.04809 & 0.168523 & 0.224697 & P1L & fusion\\
\end{longtable}
\normalsize
\end{landscape}

\begin{landscape}
\scriptsize
\setlength{\tabcolsep}{2pt}
\begin{longtable}{@{}>{\raggedright\arraybackslash}p{0.18\linewidth}>{\raggedright\arraybackslash}p{0.2\linewidth}>{\raggedright\arraybackslash}p{0.12\linewidth}>{\raggedright\arraybackslash}p{0.12\linewidth}>{\raggedright\arraybackslash}p{0.12\linewidth}@{}}
\caption{Pairwise comparisons of consistency with false discovery rate adjustment.}\label{tab:S4}\\
\toprule
\textbf{Dataset} & \textbf{LLM} & \textbf{P\_value} & \textbf{Scenario} & \textbf{Prompt}\\
\midrule
\endfirsthead
\multicolumn{5}{l}{\tablename\ \thetable{} - continued from previous page}\\
\toprule
\textbf{Dataset} & \textbf{LLM} & \textbf{P\_value} & \textbf{Scenario} & \textbf{Prompt}\\
\midrule
\endhead
\midrule
\multicolumn{5}{r}{Continued on next page}\\
\endfoot
\bottomrule
\endlastfoot
Long & GPT-O4mini & 0.001662 & P1L & fusion\\
Long & GPT-4.1 & 0.001662 & P1L & fusion\\
Long & GPT-5 & 0.001662 & P1L & fusion\\
Long & Gemini2Flash & 0.001662 & P1L & fusion\\
Long & Gemini2.5Flash & 0.002105 & P1L & fusion\\
Long & DeepSeek\_v3 & 0.001662 & P1L & fusion\\
Long & DeepSeek\_r1 & 0.001662 & P1L & fusion\\
Short & GPT-O4mini & 0.001662 & P1S & fusion\\
\end{longtable}
\normalsize
\end{landscape}

\begin{landscape}
\scriptsize
\setlength{\tabcolsep}{2pt}
\begin{longtable}{@{}>{\raggedright\arraybackslash}p{0.16\linewidth}>{\raggedright\arraybackslash}p{0.11\linewidth}>{\raggedright\arraybackslash}p{0.11\linewidth}>{\raggedright\arraybackslash}p{0.15\linewidth}>{\raggedright\arraybackslash}p{0.16\linewidth}>{\raggedright\arraybackslash}p{0.15\linewidth}>{\raggedright\arraybackslash}p{0.11\linewidth}@{}}
\caption{Augmentation: descriptive results by prompt family and scenario.}\label{tab:S5}\\
\toprule
\textbf{Model} & \textbf{Patient} & \textbf{Dataset} & \textbf{Prompt} & \textbf{LYGPC} & \textbf{$\Delta$LYGPC} & \textbf{Selectivity}\\
\midrule
\endfirsthead
\multicolumn{7}{l}{\tablename\ \thetable{} - continued from previous page}\\
\toprule
\textbf{Model} & \textbf{Patient} & \textbf{Dataset} & \textbf{Prompt} & \textbf{LYGPC} & \textbf{$\Delta$LYGPC} & \textbf{Selectivity}\\
\midrule
\endhead
\midrule
\multicolumn{7}{r}{Continued on next page}\\
\endfoot
\bottomrule
\endlastfoot
DeepSeek\_r1 & P1 & Long & A & 1.13413 & 0.026099 & 9.766667\\
DeepSeek\_r1 & P1 & Short & A & 1.056691 & -0.30763 & 8.966667\\
DeepSeek\_v3 & P1 & Long & A & 1.29763 & 0.1896 & 7.033333\\
DeepSeek\_v3 & P1 & Short & A & 1.172512 & -0.19181 & 7.033333\\
gemini\_2 & P1 & Long & A & 1.109649 & 0.001619 & 11.4\\
gemini\_2 & P1 & Short & A & 1.55625 & 0.191931 & 8\\
gemini\_25\_F & P1 & Long & A & 1.120633 & 0.012602 & 11.06667\\
gemini\_25\_F & P1 & Short & A & 1.127201 & -0.23712 & 10.6\\
\end{longtable}
\normalsize
\end{landscape}

\begin{landscape}
\scriptsize
\setlength{\tabcolsep}{2pt}
\begin{longtable}{@{}>{\raggedright\arraybackslash}p{0.16\linewidth}>{\raggedright\arraybackslash}p{0.13\linewidth}>{\raggedright\arraybackslash}p{0.13\linewidth}>{\raggedright\arraybackslash}p{0.1\linewidth}>{\raggedright\arraybackslash}p{0.1\linewidth}>{\raggedright\arraybackslash}p{0.1\linewidth}>{\raggedright\arraybackslash}p{0.1\linewidth}>{\raggedright\arraybackslash}p{0.1\linewidth}@{}}
\caption{Augmentation: pairwise comparisons of life years gained per prioritized choice, with false discovery rate adjustment.}\label{tab:S6}\\
\toprule
\textbf{Model\_A} & \textbf{Model\_B} & \textbf{d\_hat} & \textbf{P\_raw} & \textbf{P\_adj\_BH} & \textbf{Prompt} & \textbf{dataset} & \textbf{Patient}\\
\midrule
\endfirsthead
\multicolumn{8}{l}{\tablename\ \thetable{} - continued from previous page}\\
\toprule
\textbf{Model\_A} & \textbf{Model\_B} & \textbf{d\_hat} & \textbf{P\_raw} & \textbf{P\_adj\_BH} & \textbf{Prompt} & \textbf{dataset} & \textbf{Patient}\\
\midrule
\endhead
\midrule
\multicolumn{8}{r}{Continued on next page}\\
\endfoot
\bottomrule
\endlastfoot
DeepSeek\_r1 & Physicians & 0.026099 & 0.669311 & 0.937036 & A & Long & P1\\
DeepSeek\_v3 & Physicians & 0.1896 & 0.00348 & 0.012178 & A & Long & P1\\
gemini\_2 & Physicians & 0.001619 & 0.977971 & 0.977971 & A & Long & P1\\
gemini\_25\_F & Physicians & 0.012602 & 0.831395 & 0.969961 & A & Long & P1\\
gpt\_41 & Physicians & 0.269789 & 1.11E-05 & 7.74E-05 & A & Long & P1\\
gpt\_5 & Physicians & -0.04513 & 0.485688 & 0.849955 & A & Long & P1\\
gpt\_o4min & Physicians & -0.09054 & 0.165984 & 0.387295 & A & Long & P1\\
DeepSeek\_r1 & Physicians & 0.035944 & 0.551348 & 0.643239 & B & Long & P1\\
\end{longtable}
\normalsize
\end{landscape}

\end{document}